\documentclass[a4paper,11pt]{article}
\pdfoutput=1 % if your are submitting a pdflatex (i.e. if you have images in pdf, png or jpg format)
\usepackage{jheppub} 
\usepackage[T1]{fontenc} % if needed
\usepackage[utf8]{inputenc}
\usepackage{scalerel}

\title{\boldmath Dark Monopoles in Grand Unified Theories}

\author{Maria de Lourdes Z. P. Deglmann, }
\author{Marco A. C. Kneipp}
\affiliation{Universidade Federal de Santa Catarina (UFSC), Departamento de Física, CFM, \\ Campus Universitário, Trindade, 88040-900, Florianópolis, Brazil}

\emailAdd{m.l.deglmann@posgrad.ufsc.br}
\emailAdd{marco.kneipp@ufsc.br}

\abstract{We consider a Yang-Mills-Higgs theory with gauge group $G=SU(n)$ broken to $G_{v} = [SU(p)\times SU(n-p)\times U(1)]/Z$ by a Higgs field in the adjoint representation. We obtain monopole solutions whose magnetic field is not in the Cartan Subalgebra. Since their magnetic field vanishes in the direction of the generator of the $U(1)_{em}$ electromagnetic group, we call them Dark Monopoles. These Dark Monopoles must exist in some Grand Unified Theories (GUTs) without the need to introduce a dark sector. We analyze the particular case of $SU(5)$ GUT, where we obtain that their mass is $M = 4\pi v\,\widetilde{E}(\lambda/e^{2})/e$, where $\widetilde{E}(\lambda/e^{2})$ is a monotonically increasing function of $\lambda/e^{2}$ with $\widetilde{E}(0)=1.294$ and $\widetilde{E}(\infty)=3.262.$ We also give a geometrical interpretation to their non-abelian magnetic charge.}

\begin{document}
\maketitle \flushbottom

\section{Introduction}

There are many motivations to believe that the Standard Model is embedded in a Grand Unified Theory (GUT). There are various different candidates for such a theory, usually with several stages of symmetry breaking. One of the consequences of these GUTs is that they have topological magnetic monopoles. The 't Hooft-Polyakov monopole \cite{'tHooft,Polyakov} was the first example of such a  topological monopole for the $SO(3)$ Georgi-Glashow
model. Since then, there have been many generalizations for these monopoles, for theories with larger gauge groups $G$. In many of these theories, there is a Higgs field in the adjoint representation, which can produce a symmetry breaking of the form \cite{CorriganOlive76,GO78,GO81} $G\rightarrow G_{v} = "U(1)\times K"$, with a compact $U(1)$, which allows for the existence of topological monopoles. In general, these monopoles have magnetic charge in the abelian subalgebra of the unbroken group $G_{v}$, which can give rise to a non-vanishing magnetic charge for the electromagnetic $U(1)_{em}$ gauge group. The $SU(5)$ Grand Unified Theory is one example of such a theory, with monopoles \cite{DokosTomaras} associated to a spontaneous symmetry breaking by a Higgs field in the adjoint representation. 
In this work we shall construct monopole solutions with vanishing abelian magnetic charge. This implies that our monopoles do not interact with the $U(1)_{em}$ electromagnetic field and, therefore, we shall call them Dark Monopoles. Moreover, it is well-known that the nature of Dark Matter is one of the biggest open problems in physics. In the last decades, many candidates have been proposed (see, for instance, \cite{Gelmini,Freese} and references therein) in a variety of distinct theories. Magnetic monopoles happen to be one of these candidates \cite{SanchezHoldom,KhozeRo,BaekKo,Evslin,SatoTakahashi,LazaridesShafi,MurayamaShu}, usually associated to a dark (or
hidden) sector coupled to the Standard Model. But, since our Dark Monopoles do not have a $U(1)_{em}$ electromagnetic field, we need not to introduce a dark sector. This is an interesting feature, since we can have these monopoles contributing to dark matter in the standard Grand Unified Theories. And, even if they do not have a relevant contribution to Dark Matter (due to inflation), they are still an interesting solution since they are a new type of  monopole which must exist in Grand Unified Theories with Higgs field in the adjoint representation.

Monopoles with a magnetic flux in a non-abelian direction have been constructed for a Yang-Mills-Higgs theory with $G=SU(3)$ broken to $ "SU(2) \times U(1)" $ \cite{CorriganOliveFairlieNuyts} (see also \cite{GO78,Burzlaff,KunzMasak}). They were associated to the $su(2)$ subalgebra generated by the Gell-Mann matrices $\lambda_{2},\,\lambda_{5}$ and $\lambda_{7}$ and an ansatz was constructed using some general arguments of symmetry.
On the other hand, in the present work we consider a Yang-Mills-Higgs theory with an arbitrary gauge group $SU(n)$ broken to 
\begin{equation*}
G_{v}=[SU(p)\times SU(n-p)\times U(1)]/Z\,,
\end{equation*}
by a scalar field in the adjoint representation and we use a general procedure \cite{Coleman50,WeinbergLondonRosner} to construct the monopole asymptotic configuration, associated to some $su(2)$ subalgebras. We consider $su(2)$ subalgebras with generators $M_{a}$, which are linear combinations
of some step operators. Thus, the asymptotic form of the gauge and magnetic fields are linear combinations of the generators $M_{a}$, while the asymptotic form of the scalar field is a linear combination of generators $S$ and $Q_{a},\,a=0,\,\pm1,\,\pm2\,,$ which form, respectively, a singlet and a quintuplet under the $su(2)$ subalgebra. 

From these asymptotic configurations, we construct an ansatz for the whole space and calculate the Hamiltonian. Then, we obtain the second order differential equations for the profile
functions.  Addicionally, we obtain  the numerical solution for these equations in the case $G=SU(5)$, for some particular coupling constant values.
Moreover, we show  that the mass of a Dark Monopole is a monotonically increasing function of $\lambda/e^{2}$, and for $G=SU(5)$, the mass range at the classical level is 
\begin{equation*}
M=\frac{4\pi v}{e}\,\widetilde{E}(\lambda/e^{2})
\end{equation*}
where $\widetilde{E}(0)=1.294$ and $\widetilde{E}(\infty)=3.262$. It is interesting to note that due to the fact that for the Dark Monopoles $B_{i}$ and $D_{i}\phi$ are linear combinations of different generators,
the Bogomolny equation $B_{i}=D_{i}\phi$ does not have a non-trivial solution. 

We also construct a Killing vector $\zeta$ associated to an asymptotic symmetry of the Dark Monopole and show that these monopoles have a conserved current in a non-abelian direction. The associated magnetic charge $Q_{M}$ is quantized in multiples of $8\pi/e$ and we give a geometrical interpretation to this charge. Although Dark Monopoles
are associated to the trivial sector of $\Pi_{1}(G_{v})$, the conservation of $Q_{M}$ could prevent them to decay. Our construction is quite general and, in principle, it could be generalized to other gauge groups.

This paper is organized as follows: in section \ref{General_Construction} we review a general procedure to construct the asymptotic configuration for the fields of a monopole. Then, in section \ref{DM_Construction} we show the specific construction of the asymptotic configuration of a Dark Monopole for the gauge group $SU(n)$ and we propose the ansatz. We also show that our solution is not equivalent to any other solution whose magnetic field lies in the Cartan subalgebra. In section \ref{Hamiltonian_and_EoMs} we get the Hamiltonian for our Dark Monopoles  and the radial equations for the profile functions. We also obtain the numerical solution for these equations, for some particular coupling constant values, and the mass range for the $SU(5)$ Dark  Monopole. Finally, in section \ref{Magnetic_Charge} we construct a Killing vector associated to an asymptotic symmetry of the Dark Monopole and the corresponding current and conserved charge. We conclude with a summary of the results and with a discussion on the possible cosmological implications of Dark Monopoles.

\section{Magnetic monopoles in non-Abelian theories}\label{General_Construction}

In this section we will fix some conventions and review a general
construction of the asymptotic form of monopole solutions. We will
consider a Yang-Mills-Higgs theory in $3+1$ dimensions with gauge
group $G$ of rank $r$, which is simple and simply connected, and
with a real scalar field $\phi=\phi_{a}T_{a}$ in the adjoint representation.
The generators $T_{a}$ form an orthogonal basis for the Lie algebra
$g$ of $G$ which satisfy $Tr\left(T_{a}T_{b}\right)=y\,\delta_{ab}$,
where $\psi^{2}y$ is the Dynkin index of the representation and $\psi$
is the highest root of $g$. We will also use the Cartan-Weyl basis
with Cartan elements $H_{i}$, which form a basis for the Cartan subalgebra
$\mathcal{H}$, and step operators $E_{\alpha}$, satisfying the commutation
relations 
\begin{align}
\left[H_{i},H_{j}\right] & =0\,,\nonumber \\
\left[H_{i},E_{\alpha}\right] & =\alpha^{(i)}E_{\alpha}\,,\label{eq:commutators Lie algebra}\\
\left[E_{\alpha},E_{\beta}\right] & =\begin{cases}
N_{\alpha,\beta}\,E_{\alpha+\beta}\,\text{ if \ensuremath{\alpha+\beta} is a root,}\\
\frac{2\alpha\cdot H}{\alpha^{2}}\,\text{ if \ensuremath{\alpha=-\beta},}\\
0,\text{ otherwise.}
\end{cases}\nonumber 
\end{align}
Moreover, 
\begin{align}
\text{Tr}\left(H_{i}H_{j}\right) & =y\,\delta_{ij}\,,\nonumber \\
\text{Tr}\left(E_{\alpha}E_{\beta}\right) & =y\,\frac{2}{\alpha^{2}}\delta_{\alpha,-\beta}\,,\label{eq:traces Lie algebra}\\
\text{Tr}\left(H_{i}E_{\alpha}\right) & =0\,.\nonumber 
\end{align}
For an arbitrary root $\alpha$ we define the generators 
\begin{align}
T_{1}^{\alpha} & \equiv\frac{E_{\alpha}+E_{-\alpha}}{2}\,,\nonumber \\
T_{2}^{\alpha} & \equiv\frac{E_{\alpha}-E_{-\alpha}}{2i}\,,\label{eq:su(2) subalgebras}\\
T_{3}^{\alpha} & \equiv\frac{\alpha\cdot H}{\alpha^{2}}\,,\nonumber 
\end{align}
which form an $su(2)$ subalgebra. We will denote by $\alpha_{i}$,
$i=1,\,2,\,\dots,\,r$, the simple roots, and by $\lambda_{i}$, $i=1,\,2,\,\dots,\,r$
the fundamental weights of $g$, which satisfy the relation 
\begin{align*}
\frac{2\alpha_{i}\cdot\lambda_{j}}{\alpha_{i}^{2}}=\delta_{ij}\,.
\end{align*}

Let $\phi_{0}$ be the vacuum configuration of the theory which spontaneously
breaks the gauge group $G$ to $G_{v}$. By a gauge transformation,
the vacuum configuration $\phi_{0}$, can be made to lie in the Cartan
subalgebra $\mathcal{H}$, that is $\phi_{0}=u\cdot H,$ where $u$
is a vector. For a vacuum in the adjoint representation, all the generators
of $G_{v}$ must commute with $\phi_{0}$ and form a Lie algebra which
we will call $g_{v}$. Since $\phi_{0}$ commutes with itself and
all other generators of $G_{v}$, it will generate an invariant subgroup
$U(1)$ of $G_{v}$. In order for this $U(1)$ to be compact, the vector
$u$ must be proportional to a fundamental weight of $G$ \cite{GO81}.
Then, in this case, the symmetry breaking by $\phi_{0}$ in the adjoint
representation, $G_{v}$ will have the general form \cite{CorriganOlive76,GO81}
\begin{equation}
G_{v}=\frac{K\times U(1)}{Z}\,,\label{eq:G_0}
\end{equation}
where $K$ is a semisimple group, $Z$ is a discrete subgroup of the center of $K$, $Z(K)$, which belongs to $U(1)$ and $K$, i.e., $Z=U(1)\cap K$. We shall call this a minimal symmetry breaking. Then, from the condition 
\[
D_{i}\phi_{0}=0\,,
\]
that the vacuum $\phi_{0}$ fulfills, we obtain that the vacuum manifold
$G/G_{v}$ satisfies \cite{GO78}, 
\[
\Pi_{2}(G/G_{v})\cong\Pi_{1}(G_{v})\cong\mathbb{Z}\,.
\]

For a static configuration with $W_{0}=0$, $D_{0}\phi=0$ and $G_{i0}=0$,
the energy is 
\begin{equation}
E=\int d^{3}x\left\{ \frac{1}{2y}\,\text{Tr}\left(B_{i}B_{i}\right)+\frac{1}{2y}\,\text{Tr}\left(D_{i}\phi D_{i}\phi\right)+V(\phi)\right\} \,,\label{eq:Hamiltonian_Initial}
\end{equation}
where we define 
\begin{align*}
D_{\mu}\phi & =\partial_{\mu}\phi+ie\,[W_{\mu},\phi]\,,\\
V(\phi) & =\frac{\lambda}{4}\left(\frac{\text{Tr}(\phi\phi)}{y}-v^{2}\right)^{2}\,.
\end{align*}
In order for the monopole solution to have finite energy, at $r\to\infty$,
\begin{align}
V(\phi) & =0\,,\nonumber \\
D_{i}\phi & =0\,,\label{eq:vacuum conditions}\\
B_{i} & =0\,.\nonumber 
\end{align}
The first condition implies that asymptotically, $\phi$ must lay in the vacuum manifold. We can then consider that, asymptotically, $\phi$ is a gauge transformation of $\phi_{0}$, that is, 
\begin{equation}
\phi(r\to\infty,\theta,\varphi)=g(\theta,\varphi)\,\phi_{0}\,g(\theta,\varphi)^{-1}\,.\label{eq:asymptotic conf. scalar field}
\end{equation}
By similar arguments the gauge field has the form \cite{WeinbergLondonRosner,Coleman50}
\begin{equation}
W_{i}(r\to\infty,\theta,\varphi)=g(\theta,\varphi)\,W_{i}^{(0)}\,g(\theta,\varphi)^{-1}+\frac{i}{e}\,\left(\partial_{i}g(\theta,\varphi)\right)\,g(\theta,\varphi)^{-1}\,,\label{eq:general asymptotic gauge field}
\end{equation}
where 
\begin{align}
W_{r}^{(0)} & =0=W_{\theta}^{(0)}\,,\label{string-gauge-1}\\
W_{\phi}^{(0)} & =\frac{1-\cos\theta}{e}\,M_{3}\label{string-gauge-2}\,,
\end{align}
and $M_{3}$ is a generator of $g_{v}$, in order for $D_{i}\phi=0$.
Let us consider that there exist two other generators, $M_{1}$ and
$M_{2}$ of $g$, which do not belong to $g_{v}$, and which together
with $M_{3}$ form a $su(2)$ algebra 
\[
\left[M_{i},M_{j}\right]=i\,\epsilon_{ijk}\,M_{k}\,.
\]
We will call $M_{i}$ by the monopole generators. Then, in order to
remove the Dirac string singularity from $W_{\mu}^{(0)}$ in the string-gauge
and for the configuration to be spherically symmetric, we will consider
that 
\begin{equation}
g(\theta,\varphi)=\exp\left(-i\varphi M_{3}\right)\exp\left(-i\theta M_{2}\right)\exp\left(i\varphi M_{3}\right)\,.\label{eq:group element}
\end{equation}
The asymptotic gauge field \eqref{eq:general asymptotic gauge field}
can be written in Cartesian coordinates as 
\begin{equation}
W_{i}(r\to\infty)=-\epsilon_{ijk}\,\frac{n^{j}}{er}\,M_{k},\label{eq:asymptotic gauge field in Cartesian coord}
\end{equation}
with $n^{j}=x^{j}/r$. The gauge field configuration gives rise to the asymptotic
magnetic monopole field 
\begin{equation}
B_{i}(r\to\infty)=-\frac{n^{i}}{er^{2}}\,n^{a}M_{a}=-\frac{x^{i}}{er^{3}}\,gM_{3}g^{-1}\,.\label{eq:asymptotic magnetic field}
\end{equation}
The group element \eqref{eq:group element} is single-valued, except
at $\theta=\pi$, where \cite{WeinbergLondonRosner} 
\begin{equation}
g(\pi,\varphi)g(\pi,0)^{-1}=\exp(-2i\varphi M_{3})=h(\varphi)\,.\label{eq:closed loop}
\end{equation}
Since $M_{3}$ is a $su(2)$ generator, it has integer or half-integer
eigenvalues, and therefore $h(\varphi)$, $0\leq\varphi\leq2\pi$,
provides a closed loop in $G_{v}$ which is associated to sectors
of $\Pi_{1}(G_{v})$ and the monopole solutions are associated to
these topological sectors.

For the symmetry breaking $G\rightarrow G_{v}$, with $G_{v}$ given
by \eqref{eq:G_0}, we can recover the asymptotic form of the 't Hooft-Polyakov
monopole \cite{'tHooft,Polyakov} and generalizations to larger gauge groups \cite{Bais78,Weinberg82}, considering the $su(2)$ subalgebras formed by the generators  $M_{i}=T_{i}^{\alpha}$, for roots $\alpha$ such
that $\alpha\cdot u\neq0$, for $\phi_{0}=u\cdot H$. Since $T_{3}^{\alpha}\in\mathcal{H}$,
the magnetic field \eqref{eq:asymptotic magnetic field} for these
monopoles is in the Cartan subalgebra $\mathcal{H}$, up to conjugation
by $g(\theta,\varphi)$.

\section{The Dark Monopoles}\label{DM_Construction}

Now we want to construct monopoles with asymptotic magnetic field which is not in the Cartan subalgebra ${\cal H}$, that is $M_{3}\notin\mathcal{H}$,  in theories with $\phi$ in the adjoint representation, which are relevant to some GUTs. We will call them Dark Monopoles, since their magnetic field vanishes in the direction of the generator of the electromagnetic group $U(1)_{em}$, which we consider to be in ${\cal H}$.
Since $[M_{3}, \phi_{0}]=0$ and $M_{3}$ is hermitian, it is usually considered that $M_{3}$ belongs to the same Cartan subalgebra as $\phi_{0}$. However, this is not necessary when $G_{v}$ is a non-abelian gauge group. In fact, more monopole solutions can be obtained if we do not impose this condition. A nice analysis of this problem in the $SU(3)\to U(2)$ case can be found in \cite{BaisSchroers1,BaisSchroers2}. In the case of $\mathbf{Z}_{2}$ monopoles, for theories where $\phi$ is not in the adjoint representation, one can have solutions with $M_{3}$ in the direction of some step operators \cite{WeinbergLondonRosner,KneippLiebgott1,KneippLiebgott2}. Also note that string-vortex solutions with magnetic fields as 
combinations of step operators have been constructed for Yang-Mills-Higgs theories for various gauge groups \cite{AryalEverett,Ma,VilenkinShellard,HindmarshKibbble,DavisDavis,KibbleLozano,KneippLiebgott3}. For simplicity, we will
consider that the gauge group is $G=SU(n)$ and that 
\begin{equation}
\phi_{0}=v\,\frac{\lambda_{p}\cdot H}{|\lambda_{p}|}\,,\label{eq: scalar vacuum}
\end{equation}
where $\lambda_{p}$ is an arbitrary fundamental weight of $su(n)$.
This vacuum, spontaneously breaks $SU(n)$ to \cite{GO81} 
\begin{equation*}
G_{v}=[SU(p)\times SU(n-p)\times U(1)]/Z\,.
\end{equation*}
It is useful to recall that the roots of the algebra $su(n)$ in the basis of the simple roots have the form 
\begin{equation}
\pm\left(\alpha_{i}+\alpha_{i+1}+\cdots+\alpha_{j-2}+\alpha_{j-1}\right)=\pm\left(e_{i}-e_{j}\right),\,\,1\leq i<j\leq n\,,\label{eq:roots su(n)}
\end{equation}
where $e_{i}$ are orthonormal vectors in a $n$-dimensional vector space, and therefore the roots of $su(n)$ have the same length square, which is equal to 2. A root $(e_{i}-e_{j})$ is positive if $i<j$, is negative
if $i>j$ and is a simple root if $j=i+1$. A simple way to obtain the commutators between the step operators $E_{\alpha}$ in an arbitrary representation of $su(n)$ is to use the fact that in the $n$-dimensional representation of $su(n)$, the step operator $E_{\alpha}$ associated
to the root $\alpha=(e_{i}-e_{j})$, is represented by the $n\times n$ matrix $(E_{ij})_{kl}=\delta_{ik}\delta_{jl}$ and that the commutator of two generators is the same in any representation.

Let us also recall that for the Cartan involution of an arbitrary
semisimple Lie algebra $g$ \cite{Helgason}, 
\begin{align*}
\sigma(H_{i}) & =-H_{i},\\
\sigma(E_{\alpha}) & =-E_{-\alpha}\,,
\end{align*}
and $g$ can be decomposed as 
\[
g=g^{(0)}\oplus g^{(1)}
\]
where 
\begin{align}
g^{(0)} & =\left\{ \left(E_{\alpha}-E_{-\alpha}\right)/2i,\text{ for \ensuremath{\alpha>0}}\right\} \,,\label{eq:grades}\\
g^{(1)} & =\left\{ H_{i},\,i=1,\,2,\,\dots,\,r;\,\left(E_{\alpha}+E_{-\alpha}\right)/2,\,\text{ for \ensuremath{\alpha>0}}\right\} \,.\nonumber 
\end{align}
Then $g^{(0)}$ forms a subalgebra of $g$ and the generators of $g^{(1)}$
form a representation of $g^{(0)}.$ For example, for $g=su(3)$,
there are three generators in $g^{(0)}$ which form a $su(2)$ subalgebra
and there are five generators in $g^{(1)}$ which form a quintuplet
of this $su(2)$ subalgebra.

In order to construct Dark Monopole solutions, we shall consider that
the monopole generators $M_{i}$, which form a $su(2)$ subalgebra,
belong to $g^{(0)}$. Then, $\phi_{0}$, which is in $g^{(1)}$, will
be in a representation of this $su(2)$, as we will see later on. Using
the definition of eq.\eqref{eq:su(2) subalgebras}, we will consider
that $M_{3}=2T_{2}^{\alpha}$, $M_{1}=2T_{2}^{\beta}$ and $M_{2}=2T_{2}^{\gamma}$
where $\alpha,\,\beta,\,\gamma$ are roots of $su(n).$ Then, the
condition that they form a $su(2)$ algebra implies that $\alpha+\beta+\gamma=0$.
Now, since $M_{3}\in g_{v}$, then $\left[M_{3},\phi_{0}\right]=0$,
which implies that $\alpha\cdot\lambda_{p}=0$, and therefore $\alpha$
does not have the simple root $\alpha_{p}$ in its expansion in the
simple root basis. Thus, for $\alpha=e_{i}-e_{j}$, if $i<j,$ either
$i>p$ or $j\leq p$, and if $i>j$, either $i\leq p$ or $j>p$.
On the other hand, since $M_{1}$ and $M_{2}$ do not belong to $g_{v},$
then $\beta\cdot\lambda_{p}\neq0$ and $\gamma\cdot\lambda_{p}\neq0$,
which implies that $\beta$ and $\gamma$ have the simple root $\alpha_{p}$
in their expansion in the simple root basis. Then, denoting by $T_{a}^{ij},\,a=1,\,2,\,3,$
the generators defined in Eq.\eqref{eq:su(2) subalgebras} for $\alpha=e_{i}-e_{j}$,
we can conclude that the possible monopole generators, for $\alpha$
positive are 
\begin{align}
M_{3} & =2T_{2}^{ij}\,,\nonumber \\
M_{1} & =2T_{2}^{jk}\,,\label{eq:su(2) positive alpha}\\
M_{2} & =2T_{2}^{ki}\,,\nonumber 
\end{align}
where there are two possibilities: a) $1\leq i<j\leq p$ and $j<k,$
with $\,p<k\leq n$; b) $p<i<j\leq n$ and $k<j$ with $1<k\leq p$.
Each of these $su(2)$ subalgebras can be labeled by these three numbers
$i,\,j,\,k$. On the other hand, when $\alpha$ is a negative root,
$i>j$, which can be seen as an exchange between $i\leftrightarrow j$
in the cases above. We should also remark that there may be other $su(2)$ subalgebras, with $M_3$ being a combination of step operators, from which we could construct other Dark Monopole solutions. However, for simplicity, in this work we will only consider the $su(2)$ subalgebras related to positive roots, given by eq.\eqref{eq:su(2) positive alpha}.

Note that each set of $M_{i},\,i=1,2,3$, generates an $SO(3)$ subgroup
of $SU(n)$\footnote{For $G=SU(3)$, in the three dimensional representation, these generators
correspond to the Gell-Mann matrices $\lambda_{7},\,-\lambda_{5},\,\lambda_{2}$. }. However, the associated closed loop $h(\varphi),\,0\leq\varphi\leq2\pi$, 
given by eq. \eqref{eq:closed loop}, is contractible. Therefore, these monopoles are associated to the
trivial topological sector of $\Pi_{1}(G_{v})$.

For each $su(2)$ subalgebra, we can construct a monopole solution. And in order to obtain the asymptotic configuration
of the scalar field \eqref{eq:asymptotic conf. scalar field} for each of them, it is convenient to decompose $\phi_{0}$ as 
\begin{equation}
\phi_{0}=v\,\left(S+\frac{2Q_{0}}{\sqrt{6}\,|\lambda_{p}|}\right)\,,\label{eq:vacuum decomposition}
\end{equation}
where 
\begin{align*}
Q_{0} & =\frac{2}{\sqrt{6}}\left(T_{3}^{ik}+T_{3}^{jk}\right)\,,\\
S & =\frac{\lambda_{p}\cdot H}{|\lambda_{p}|}-\frac{2Q_{0}}{\sqrt{6}\,|\lambda_{p}|}\,,
\end{align*}
with $\text{Tr}(Q_{0}Q_{0})=y$ and 
\[
\left[M_{3},Q_{0}\right]=0=\left[M_{3},S\right]\,.
\]
Moreover, $[M_{\pm},S]=0$, where $M_{\pm}=M_{1}\pm iM_{2}$. Therefore,
$S$ is a singlet. On the other hand, one can check that $Q_{0}$
belongs to a quintuplet together with the generators 
\begin{align*}
Q_{\pm1} & =\pm\left(T_{1}^{ik}\pm i\,T_{1}^{jk}\right)\,,\\
Q_{\pm2} & =-\left(T_{3}^{ij}\pm i\,T_{1}^{ij}\right)\,,
\end{align*}
satisfying the commutation relations 
\begin{align}
\left[M_{3},Q_{m}\right] & =m\,Q_{m}\,,\label{eq:commutators of M and Q,a}\\
\left[M_{\pm},Q_{m}\right] & =c_{l,m}^{\pm}\,Q_{m\pm1}\,,\label{eq:commutators of M and Q,b}
\end{align}
where $c_{l,m}^{\pm}=\sqrt{l(l+1)-m(m\pm1)}$ with $l=2$.

Although for any $su(2)$ subalgebra $M_{i}$, the generators $Q_{m}$ always form a quintuplet and therefore $l=2$, we will continue to write $l$ to keep track of this constant. It can also be useful for possible generalizations of the Dark Monopole construction with different $l$ for other gauge groups.

Since $M_{i}\in g^{(0)}$ and $Q_{m}\in g^{(1)}$, then, 
\[
\text{Tr}\left(M_{i}Q_{m}\right)=0\,.
\]
Moreover, since 
\[
\text{Tr}(Q_{m}[Q_{p},M_{3}])=\text{Tr}(Q_{p}[M_{3},Q_{m}])\,,
\]
it results that $\text{Tr}(Q_{m}Q_{p})=0$ if $p\neq-m$. Similarly,
from $\text{Tr}\left(Q_{m}\left[Q_{-(m+1)},M_{\pm}\right]\right)$,
results that 
\[
\text{Tr}\left(Q_{m}Q_{-m}\right)=-\text{Tr}\left(Q_{m+1}Q_{-(m+1)}\right)\,.
\]
Therefore, we can conclude that 
\begin{equation}
\text{Tr}\left(Q_{m}Q_{p}\right)=(-1)^{m}\,y\,\delta_{m,-p}\,.\label{Trace_for_Q_m}
\end{equation}
Finally, from the definition of the generators $M_{i}$, it results
that \begin{subequations} 
\begin{align}
\text{Tr}(M_{i}M_{j}) & =2\,y\,\delta_{ij}\,,\label{Trace_for_Ms}\\
\text{Tr}(M_{+}M_{-}) & =4\,y\,.
\end{align}
\end{subequations}

Now, since $Q_{m}\in g^{(1)}$, then $[Q_{m},Q_{p}]\in g^{(0)}$.
Thus, 
\[
\left[Q_{m},Q_{p}\right]=A_{mp}\,M_{3}+B_{mp}^{+}\,M_{+}+B_{mp}^{-}\,M_{-}+\sum_{\delta}D_{mp}^{\delta}\,T_{2}^{\delta}\,,
\]
where $A_{mp},\,B_{mp}^{\pm},\,D_{mp}^{\delta}$ are constants and
$T_{2}^{\delta}$ are other possible generators of $g^{(0)}$. Then,
taking the trace of this commutator with $M_{3}$, $M_{\pm}$ and
$T_{2}^{-\delta}$, and using the previous results, we can conclude
that 
\[
\left[Q_{m},Q_{p}\right]=(-1)^{m}\left(\frac{m}{2}\,M_{3}\,\delta_{m,-p}-\frac{1}{4}\,c_{l,p}^{-}\,M_{+}\,\delta_{m,-p+1}-\frac{1}{4}\,c_{l,p}^{+}\,M_{-}\,\delta_{m,-(p+1)}\right)\,.
\]
This set of generators $M_{i},\,Q_{m}$ form an $su(3)$ subalgebra
of $su(n)$, since they are linear combinations of the generators
$T_{a}^{ij},\,T_{a}^{ik},\,T_{a}^{jk},$ $a=1,\,2,\,3$.

In order to construct the asymptotic form for the scalar field, let
us recall that in a $(2j+1)$ irreducible representation of a $su(2)$
algebra with generators $J_{i},\,i=1,\,2,\,3,$ and with eigenstates
$\left|j,m\right\rangle $, the spherical harmonics can be written
as \cite{WuKiTung}, 
\[
Y_{jm}^{*}(\theta,\varphi)=\sqrt{\frac{2j+1}{4\pi}}\,D_{m0}^{j}(\phi,\theta,0),
\]
where 
\[
D_{m0}^{j}(\phi,\theta,0)=\left\langle j,m\right|\exp(-i\varphi J_{3})\exp(-i\theta J_{2})\left|j,0\right\rangle =e^{-i\varphi m}\,d_{m0}^{j}(\theta)
\]
and 
\begin{align*}
d_{m0}^{j}(\theta) & =\left\langle j,m\right|\exp(-i\theta J_{2})\left|j,0\right\rangle \\
 & =\delta_{m0}+\sum_{n=1}^{\infty}\frac{\left(-i\theta\right)^{n}}{n!}\left[\left(D^{j}(J_{2})\right)^{n}\right]_{m0}\,,
\end{align*}
with $D^{j}(J_{i})_{m'm}=\left\langle j,m'\right|J_{i}\left|j,m\right\rangle $.

From Eqs. \eqref{eq:asymptotic conf. scalar field}, \eqref{eq:group element}
and \eqref{eq:vacuum decomposition}, the asymptotic form for the
scalar field can be written as 
\[
\phi(r\to\infty,\theta,\varphi)=v\,\left(S+\frac{2}{\sqrt{6}\,|\lambda_{p}|}\,g(\theta,\varphi)Q_{0}g(\theta,\varphi)^{-1}\right)\,.
\]
The commutation relations \eqref{eq:commutators of M and Q,a} and 
\eqref{eq:commutators of M and Q,b} can be written as 
\[
\left[M_{i},Q_{m}\right]=D^{l}(M_{i})_{m'm}Q_{m'}\,,
\]
where $D^{l}(M_{i})_{m'm}$ is the $(2l+1)-$dimensional representation
of the $su(2)$ generator $M_{i}$ in the basis of the $Q_{m}$'s.
Then 
\begin{align*}
\exp\left(-i\theta M_{2}\right)\,Q_{0}\,\exp\left(i\theta M_{2}\right) & =Q_{0}+\sum_{n=1}^{\infty}\frac{\left(-i\theta\right)^{n}}{n!}\underbrace{\left[M_{2},\left[M_{2},\dots,\left[M_{2},Q_{0}\right]\right]\right]}_{n}\\
 & =\sum_{m}\left\{ \delta_{m0}+\sum_{n=1}^{\infty}\frac{\left(-i\theta\right)^{n}}{n!}\left[\left(D^{l}(M_{2})\right)^{n}\right]_{m0}\right\} \,Q_{m}\\
 & =\sum_{m}d_{m0}^{l}(\theta)\,Q_{m}\,.
\end{align*}
Hence, 
\begin{align*}
g(\theta,\varphi)\,Q_{0}\,g(\theta,\varphi)^{-1} & =\sum_{m}e^{-i\varphi m}\,d_{m0}^{l}(\theta)\,Q_{m}\\
 & =\left(\frac{4\pi}{2l+1}\right)^{1/2}\sum_{m}Y_{lm}^{*}(\theta,\varphi)\,Q_{m}\,.
\end{align*}
Therefore, the asymptotic configuration for the scalar field is 
\begin{equation}
\phi(r\to\infty,\theta,\varphi)=v\,S+\alpha\sum_{m}Y_{lm}^{*}(\theta,\varphi)\,Q_{m}\,,\label{eq:DM asymptotic scalar}
\end{equation}
with 
\begin{equation}
\alpha=\frac{2v}{\sqrt{6}\,|\lambda_{p}|}\,\sqrt{\frac{4\pi}{2l+1}}\,\label{eq:alpha}
\end{equation}
and $l = 2$. From this asymptotic configuration, we can propose an ansatz for the
whole space as 
\begin{equation}
\phi(r,\theta,\varphi)=\phi_{s}+\phi_{q}(r,\theta,\varphi)\label{eq:Dark monopole scalar ansatz}
\end{equation}
with 
\begin{align}
\phi_{s} & =v\,S,\label{eq: phi_s}\\
\phi_{q}(r,\theta,\varphi) & =\alpha\,f(r)\sum_{m}Y_{lm}^{*}(\theta,\varphi)\,Q_{m}\,,\label{eq: phi_q}
\end{align}
where $f(r)$ is a radial function such that $f(r=0)=0$ and $f(r\to\infty)=1$.

From the asymptotic gauge field configuration \eqref{eq:asymptotic gauge field in Cartesian coord},
one can propose the ansatz 
\begin{equation}
W_{i}=-\frac{[1-u(r)]}{er}\epsilon_{ijk}n^{j}M_{k},\label{eq:ansatz gauge field}
\end{equation}
with the radial function $u(r)$ satisfying the conditions, $u(r=0)=1$
and $u(r\to\infty)=0$. From this gauge field we obtain the magnetic
field 
\begin{equation}
B_{i}=\left(\frac{u'}{er}\,P_{T}^{ik}+\frac{u^{2}-1}{er^{2}}\,P_{L}^{ik}\right)M_{k}\,,\label{eq:ansatz magnetic field}
\end{equation}
where $P_{T}^{ik}=\delta^{ik}-n^{i}n^{k}$, $P_{L}^{ik}=n^{i}n^{k}$
and $u'(r)$ stands for $du/dr$.

Using the fact that 
\begin{equation}
i\epsilon_{kab}\,x^{a}\,\partial_{b}Y_{lm}^{*}=D^{l}(M_{k})_{m'm}Y_{lm'}^{*}\,,\label{property_Y}
\end{equation}
it is direct to verify that our solution is spherically symmetric
with respect to 
\begin{equation}
J_{i}\equiv-i\epsilon_{ijk}\,x^{j}\,\partial_{k}+M_{i}\,,
\end{equation}
which means that \eqref{eq:Dark monopole scalar ansatz} and \eqref{eq:ansatz gauge field}
satisfies 
\begin{align}
\left[J_{i},\phi\right] & =0\,,\\
\left[J_{i},W_{j}\right] & =i\epsilon_{ijk}W_{k}\,.
\end{align}

\subsection{On the equivalence between solutions}

After constructing the ansatz for our Dark Monopoles, we must discuss the reason why our solution is not equivalent to any other solution whose magnetic field lies in the Cartan subalgebra $\mathcal{H}$. We recall that the arguments we present here are valid for monopoles in the case of minimal symmetry breaking. First, let us denote by $(\phi^{\scaleto{M}{3.5pt}}, W_{i}^{\scaleto{M}{3.5pt}})$ a field configuration at infinity in the positive $z-$direction and, therefore, $\phi^{\scaleto{M}{3.5pt}} = \phi_{0}$. At this point, the gauge field takes values in the $su(2)$ subalgebra of the generators $M_{i}$, $i=1,2,3$. 
However, note that this configuration is not unique, since we can obtain an equally valid solution by means of a global gauge transformation $P$. Under this transformation, we have that \cite{BaisSchroers1,BaisSchroers2}
\begin{align*}
\phi^{\scaleto{P}{3.5pt}} 	&= P\,\phi^{\scaleto{M}{3.5pt}}\,P^{-1},\\
W_{i}^{\scaleto{P}{3.5pt}} 	&= P\,W_{i}^{\scaleto{M}{3.5pt}}\,P^{-1}\,,
\end{align*} 
while 
\begin{equation*}
M_{i}^{\scaleto{P}{3.5pt}} = P\,M_{i}\,P^{-1}\,,
\end{equation*} 
are also generators of an $su(2)$ subalgebra of $G$. Now, since we want to preserve the symmetry breaking, i.e.,  $\phi^{\scaleto{P}{3.5pt}} = \phi^{\scaleto{M}{3.5pt}} = \phi_{0}$, we see that $P$ must belong to $G_{v}$.

However, note that since $P$ is position independent, for other directions than the positive $z-$direction this global $G_{v}$ action does not leave the Higgs field at infinity invariant. This follows from the fact that for a general direction $\hat{r}$ this $P$ is not an element of the unbroken group $G_{v}(\hat{r})$, which is position dependent. So, even if two monopole solutions are related by the conjugation of an element $P \in G_{v}$ in the north pole, this global gauge transformation cannot be implemented to the whole asymptotic configuration because $P$ will not belong to the local unbroken gauge group $G_{v}(\hat{r})$.

In fact, we can not even define a gauge transformation $P(\theta,\varphi) = g(\theta,\varphi)\,P\,g^{-1}(\theta,\varphi)$, with $g(\theta,\varphi)$ given by eq.\eqref{eq:group element}, that takes values in $G_{v}(\hat{r})$ for every direction $\hat{r}$ at infinity. This happens because the generators of $G_{v}$ which do not commute with the magnetic field, which is proportional to $M_{3}$ in the north pole, cannot be globally well-defined. This situation is the well-known problem of "Global Color" \cite{Abouelsood1,Abouelsood2,Balachandran1,Balachandran2,Balachandran3,NelsonManohar,NelsonColeman,HorvathyRawnsley1,HorvathyRawnsley2} and happens to some monopole solutions for theories with a non-abelian unbroken symmetry (NUS). 

Then, in our specific case, there are indeed global gauge transformations that take our magnetic field in the north pole to an usual one lying in $\mathcal{H}$. The simplest of such transformations is of the form $P = \exp\left(-i \pi T_{1}^{ij}/2\right)$. But from the considerations above we see that such a transformation cannot be globally implemented, which implies that our monopole solution is distinct from those with a magnetic flux in the Cartan subalgebra. We also add that there is an example \cite{Irwin} of a similar situation in the $SU(3)\to U(2)$ symmetry breaking, where two distinct monopole solutions can be related in the north pole by the global action of the $SU(2)$ subgroup of $U(2)$, while we cannot move between these solutions dynamically, implying the solutions are \emph{physically} distinct.

\section{Hamiltonian and equations of motion}\label{Hamiltonian_and_EoMs}

In this section we shall obtain the Hamiltonian for our Dark Monopole,
as well as the equations of motion (EoMs) for the profile functions.
It is important to note that the ``traditional'' BPS bound for this
monopole is zero, since $\text{Tr}(B_{i}\phi)=0$ and therefore the
magnetic charge associated to the $U(1)$ group vanishes. However,
since $B_{i}$ is a linear combination of $M_{a}$ and $D_{i}\phi$
is a linear combination of $Q_{m}$, then the Bogomolny equation \cite{Bogomolny}
$B_{i}=D_{i}\phi$ does not have a non-trivial solution. Hence, there
is no solution associated to this vanishing bound.

Let us start with the kinetic term of the scalar field. Since the
component $\phi_{s}$ is such that $\partial_{i}(\phi_{s})=0$ and
$[\phi_{s},M_{i}]=0$, it implies that $D_{i}\phi_{s}=0$. Then, from
eq.\eqref{eq:Dark monopole scalar ansatz} one can obtain that 
\begin{equation}
D_{i}\phi=\alpha\left[(\partial_{i}f)Y_{lm}^{*}+f(\partial_{i}Y_{lm}^{*})-i\frac{f(1-u)}{r^{2}}\epsilon_{ijk}x^{j}D\left(M_{k}\right)_{m'm}Y_{lm'}^{*}\right]Q_{m}\,.\label{covariant_derivative}
\end{equation}
Making use of eq.\eqref{property_Y}, eq.\eqref{covariant_derivative}
can be written as 
\begin{equation}
D_{i}\phi=\alpha\left[\frac{f'}{r}(x^{i}\,Y_{lm}^{*})+fu\,(\partial_{i}\,Y_{lm}^{*})\right]Q_{m}\,.
\end{equation}
From eq.\eqref{Trace_for_Q_m} and the fact that $Y_{lm}=(-1)^{m}\,Y_{lm}^{*}$,
one can obtain that 
\begin{equation}
\frac{1}{y}\,\text{Tr}\left(D_{i}\phi D_{i}\phi\right)=\sum_{m=-l}^{l}\alpha^{2}\left[(f')^{2}Y_{lm}Y_{lm}^{*}+f^{2}u^{2}\,\mathbf{\nabla}Y_{lm}\cdot\mathbf{\nabla}Y_{lm}^{*}\right]\,.
\end{equation}
Moreover, using the properties of Vector Spherical Harmonics (VSH)
\cite{Barrera} we obtain that 
\begin{equation}
\frac{1}{2y}\int d^{3}x\,\text{Tr}\left(D_{i}\phi D_{i}\phi\right)=4\pi\,\frac{2v^{2}}{3|\lambda_{p}|^{2}}\int_{0}^{\infty}dr\left[\frac{1}{2}r^{2}(f')^{2}+\frac{l(l+1)}{2}f^{2}u^{2}\right]\,.\label{contribution_from_Di}
\end{equation}
From the magnetic field it follows that 
\begin{equation}
\frac{1}{2y}\int d^{3}x\,\text{Tr}\left(B_{i}B_{i}\right)=4\pi\int_{0}^{\infty}dr\,\frac{1}{e^{2}r^{2}}\,[2r^{2}(u')^{2}+(1-u^{2})^{2}]\,.\label{contribution_from_B}
\end{equation}
Finally, we use eqs.\eqref{eq:Dark monopole scalar ansatz}, the
fact that 
\begin{align*}
\text{Tr}\left(SS\right) & =\left(1-\frac{2}{3\,|\lambda_{p}|}\right)y\,,\\
\text{Tr}\left(\phi_{q}\phi_{q}\right) & =\frac{2v^{2}f^{2}}{3\,|\lambda_{p}|^{2}}\text{Tr }\left(g\,Q_{0}\,g^{-1}g\,Q_{0}\,g^{-1}\right)=\frac{2v^{2}f^{2}y}{3\,|\lambda_{p}|^{2}}\,,
\end{align*}
and $\text{Tr}(SQ_{0})=0$ to obtain that 
\begin{gather}
V(\phi)=\frac{\lambda v^{4}}{9\,|\lambda_{p}|^{4}}\,(f^{2}-1)^{2}.
\end{gather}

Joining all the contributions and making the change of variables $\xi=evr$
the Hamiltonian \eqref{eq:Hamiltonian_Initial} for the Dark Monopole
will be 
\begin{align}
E & =\frac{4\pi v}{e}\int_{0}^{\infty}d\xi\left\{ \left[2(u')^{2}+\frac{(1-u^{2})^{2}}{\xi^{2}}\right]+\frac{2}{3|\lambda_{p}|^{2}}\left[\frac{1}{2}\notag\xi^{2}(f')^{2}+\frac{l(l+1)}{2}f^{2}u^{2}\right]\right.\label{full-hamiltonian}\\
 & \left.+\,\frac{\lambda}{9e^{2}|\lambda_{p}|^{4}}\,\xi^{2}(f^{2}-1)^{2}\right\} \,,
\end{align}
where $u'(\xi),\,f'(\xi)$ denote derivatives with respect to $\xi$.

The conditions for $E$ to be stationary with respect to $f(\xi)$
and $u(\xi)$ provide the equations of motion for the ansatz of the
Dark Monopole: \begin{subequations} 
\begin{align}
u'' & =\frac{l(l+1)}{6|\lambda_{p}|^{2}}\,f^{2}u+\frac{u(u^{2}-1)}{\xi^{2}}\,,\label{eom_gauge}\\
f'' & =-\frac{2}{\xi}f'+l(l+1)\,\frac{fu^{2}}{\xi^{2}}+\frac{2\lambda}{3e^{2}|\lambda_{p}|^{2}}\,f(f^{2}-1).\,\label{eom_Higgs}
\end{align}
\end{subequations} The appropriate boundary conditions for a non-singular
finite-energy solution are 
\begin{gather}
f(0)=0,\qquad u(0)=1\label{bc-0}\\
f(\xi\to\infty)=1,\qquad u(\xi\to\infty)=0.\label{bc-infty}
\end{gather}

Before looking for numerical solutions to eqs. \eqref{eom_gauge}
and \eqref{eom_Higgs}, we shall analyze the behavior of the profile
functions when $\xi\approx0$ and also when $\xi\to\infty$.

\subsection{Approximate Solutions}

When $\xi\ll1$, eq.\eqref{eom_gauge} remains non-linear, since the
dominant contribution is of the form $u''=u(u^{2}-1)/\xi^{2}$. However,
since we are looking for approximate solutions, it is reasonable to series expand \eqref{eom_gauge} about $\xi=0$ to order $\xi^{2}$. Then,
it is a trivial task to see that 
\begin{equation}
u(\xi)=1-c_{1}\,\xi^{2}\,,\label{u-near-0}
\end{equation}
with $c_{1}\in\mathbb{R}$, gives the behavior of $u(\xi)$, subject to the boundary conditions \eqref{bc-0}, near the origin. We do not bother to fix the constant $c_{1}$, since
we are only interested in the behavior of the solution.

With regard to eq.\eqref{eom_Higgs} one can see that the dominant
contribution is of the form 
\[
\xi^{2}f''+2\xi f'-l(l+1)f=0\,,
\]
where we used the approximation $u^{2}(\xi\to0)\approx1$. This equation is in
the form of the Euler-Cauchy equation. Then, the solution which satisfies
\eqref{bc-0} is 
\begin{equation}
f(\xi)=c_{2}\,\xi^{l}\,,\label{f-near-0}
\end{equation}
where $c_{2}\in\mathbb{R}$ is also an arbitrary constant. It is important to stress that solutions \eqref{u-near-0} and \eqref{f-near-0} agree with the fact that we are looking for non-singular monopole solutions. One can explicitly check that the expression of $\phi$, $W_{i}$ and $B_{i}$ are regular at the origin.

At this point, we can make an important comparison between the 't
Hooft-Polyakov monopole and our Dark Monopoles. While the behavior
of the profile function in the gauge field ansatz ($u(\xi)$) is the
same for both, in the case of the Higgs field ($f(\xi))$ we see a
distinct behavior. In the 't Hooft-Polyakov case, $f(\xi)\sim\xi$,
although in our construction $f(\xi)\sim\xi^{2}$.

Finally we analyze how the asymptotic values \eqref{bc-infty} are approached. In order to do so, it is convenient to substitute $f=(h/\xi) + 1$ in the eqs.\eqref{eom_gauge} and \eqref{eom_Higgs} and take $\xi\rightarrow\infty$,
which results in \begin{subequations} 
\begin{align}
u'' & =\frac{l(l+1)}{6|\lambda_{p}|^{2}}\,u\,,\label{asymptotic-1}\\
h'' & =\frac{4\lambda}{3e^{2}|\lambda_{p}|^{2}}\,h\,.\label{asymptotic-2}
\end{align}
\end{subequations} Thus, the solutions behave as 
\begin{align}
u(\xi) & =O\left[\exp\left(-\sqrt{\frac{l(l+1)}{6|\lambda_{p}|^{2}}}\,\xi\right)\right]\,,\\
f(\xi)-1 & =O\left[\frac{\exp\left(-\sqrt{\frac{4\lambda}{3e^{2}\lambda_{p}|^{2}}}\,\xi\right)}{\xi}\right]\,.
\end{align}
Therefore, for distances larger than the monopole core 
\[
R_{\text{core}}=\frac{1}{ev}\,\sqrt{\frac{6\,|\lambda_{p}|^{2}}{l(l+1)}}\,,
\]
the gauge field configuration \eqref{eq:ansatz gauge field} reduces
to the asymptotic form \eqref{eq:asymptotic gauge field in Cartesian coord}
and the magnetic field \eqref{eq:ansatz magnetic field} takes the
form of a hedgehog as in eq.\eqref{eq:asymptotic magnetic field}.

\begin{figure}
\centering % \begin{center}/\end{center} takes some additional vertical space
 \includegraphics[width=0.65\textwidth]{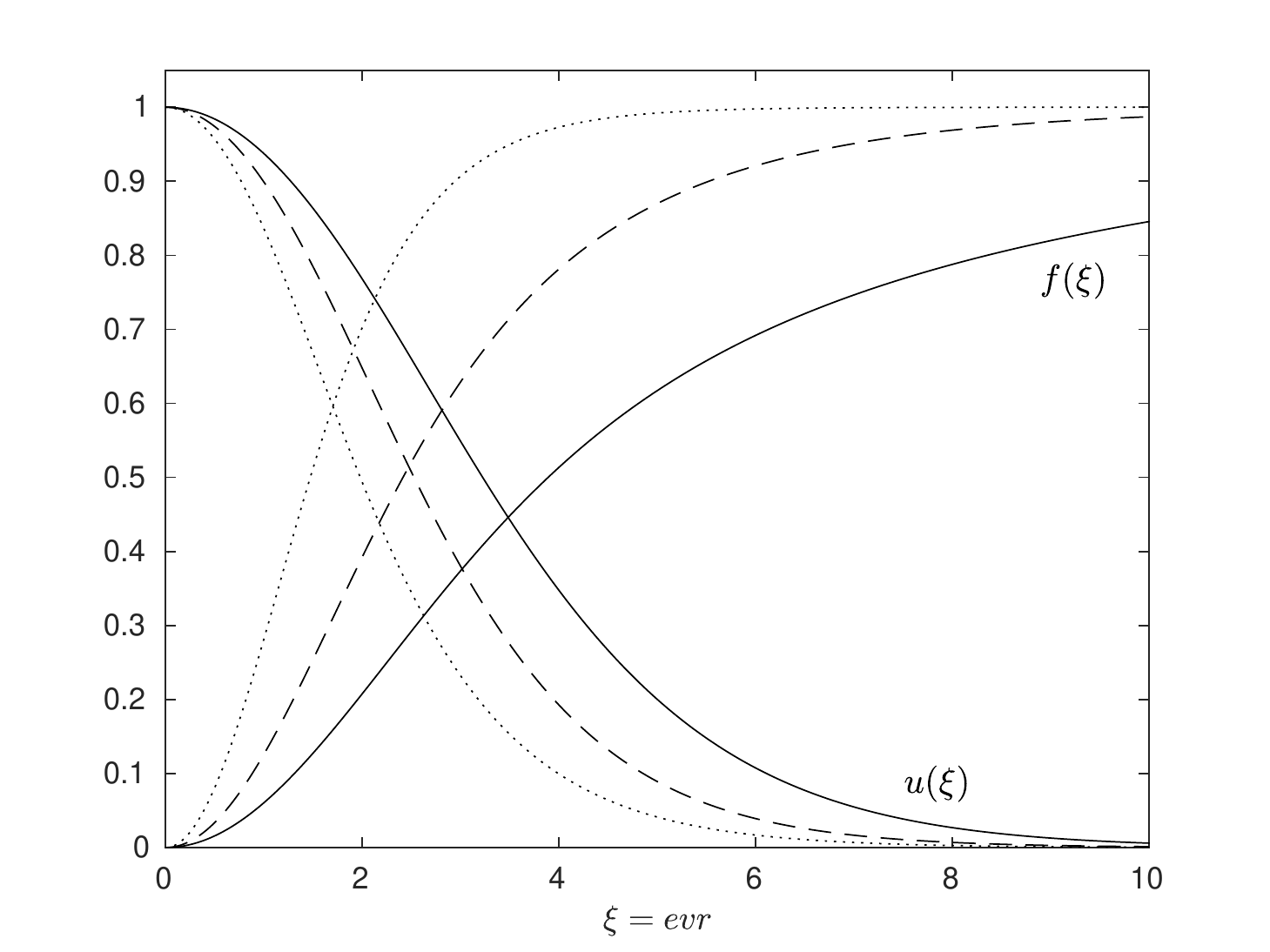}
\caption{\label{monopole-solution} The monopole profile functions $u(\xi)$
and $f(\xi)$ for $\lambda/e^{2}=0$ (solid curves), $\lambda/e^{2}=0.1$
(dashed curves) and $\lambda/e^{2}=1$ (dotted curves).}
\end{figure}

\subsection{Numerical Solution}

\begin{subequations} From the fact that we cannot find an analytical
solution to the set of equations \eqref{eom_gauge} and \eqref{eom_Higgs},
it is reasonable to look for numerical solutions. We numerically solved
the problem making use of the MATLAB\textsuperscript{{\tiny{}{}\textregistered{}}} program  \texttt{bvp4c},
which implements the solution of boundary 
value problems (BVPs). In order to do so, the system of equations 
\eqref{eom_gauge} and \eqref{eom_Higgs} were recast as a system
of first order equations of the form 
\begin{align}
u' & =v\,,\label{first-order-1}\\
v' & =\frac{l(l+1)}{6|\lambda_{p}|^{2}}\,f^{2}u+\frac{u(u^{2}-1)}{\xi^{2}}\,,\label{first-order-2}\\
f' & =w\,,\label{first-order-3}\\
w' & =-\frac{2}{\xi}w+l(l+1)\,\frac{fu^{2}}{\xi^{2}}+\frac{2\lambda}{3e^{2}|\lambda_{p}|^{2}}\,f(f^{2}-1)\,,\label{first-order-4}
\end{align}
\end{subequations} where $u,v,f\text{ and }w$ are considered to
be independent. Once more, we stress that in the case of our Dark
Monopoles $l=2$ and one can obtain several distinct solutions by
choosing different SSB patterns through the choice of $\lambda_{p}$
in the Lie algebra of $G$. These solutions must satisfy the constraints
in the behavior imposed by the approximate solutions \eqref{u-near-0}
and \eqref{f-near-0}. Figure \ref{monopole-solution} shows the solution
for the case of the SU(5) Dark Monopole, where the symmetry breaking
is of the form $SU(5)\,\rightarrow\,"SU(3)\times SU(2)\times U(1)"$,
where the quotation marks refer to the local structure of the unbroken
gauge group, only. In the SU(5) case we can take the fundamental weight
$\lambda_{p}$ to be $\lambda_{2}$ or $\lambda_{3}$, since both
of them generate the desired SSB. Then, $|\lambda_{p}|^{2}=6/5$.
One can see that this solution agrees with the expected behavior,
since $u-1\sim-\xi^{2}$ and $f\sim\xi^{2}$ near zero, while they
both reach the asymptotic values rather fast. 

The total energy of the solution, which is interpreted as the classical
mass, is given by eq.\eqref{full-hamiltonian} and to simplify the
analysis we use the rescaled mass, $\tilde{E}$, 
\[
E=M_{0}\,\tilde{E}(\lambda/e^{2}),\quad\text{where }M_{0}=\frac{4\pi v}{e}\,.
\]
Performing an analysis similar to \cite{Forgacs2005}, we can obtain
the mass range for the Dark Monopoles. Note first that $\tilde{E}$
is a monotonically increasing function of $\lambda/e^{2}$, since
\[
\frac{d\tilde{E}(\lambda/e^{2})}{d(\lambda/e^{2})}=\frac{1}{9|\lambda_{p}|^{4}}\,\int_{0}^{\infty}d\xi\,\xi^{2}(f^{2}-1)^{2}>0\,.
\]
The lower bound for the mass happens when $\lambda=0$, and numerical
integration shows that for the $SU(5)$ monopole $\tilde{E}(0)=1.294$.

Similar to the case of the 't Hooft-Polyakov monopole \cite{KirkmanZachos},
in the limit $\lambda\to\infty$ the mass of the monopole stays finite
and it is given by 
\begin{align}
E & =\frac{4\pi v}{e}\int_{0}^{\infty}d\xi\left[2(u'_{\scaleto{\infty}{2.5pt}})^{2}+\frac{(1-u_{\scaleto{\infty}{2.5pt}}^{2})^{2}}{\xi^{2}}+\frac{l(l+1)}{3|\lambda_{p}|^{2}}\,u_{\scaleto{\infty}{2.5pt}}^{2}\right]\,,\label{limit-hamiltonian}
\end{align}
since $f(\xi)\equiv1\,\forall\,\xi>0$ but $f(0)=0$. Then, the only
radial equation of motion is 
\begin{gather}
u''_{\scaleto{\infty}{2.5pt}}=\frac{l(l+1)}{6|\lambda_{p}|^{2}}\,u_{\scaleto{\infty}{2.5pt}}+\frac{u_{\scaleto{\infty}{2.5pt}}(u_{\scaleto{\infty}{2.5pt}}^{2}-1)}{\xi^{2}}\,.\label{limit-eom-gauge}
\end{gather}
Solving eq.\eqref{limit-eom-gauge} and performing the integration
in \eqref{limit-hamiltonian} gives us the upper bound for the monopole
mass. In the $SU(5)$ case, the upper bound is $\tilde{E}(\lambda\to\infty)=3.262$.
For comparison, for the 't Hooft-Polyakov monopole in the $SU(2)$
case, $\tilde{E}(\lambda=0)=1$ \cite{PrasadSommerfield} and $\tilde{E}(\lambda\to\infty)=1.787$
\cite{KirkmanZachos}.

Note that for a given SSB, where $\lambda_{p}$ is fixed, the value of the mass is the same for all the Dark Monopole solutions associated to the the $su(2)$ subalgebras \eqref{eq:su(2) positive alpha}. This follows directly from the fact that the hamiltonian is independent of the indices $i,j,k$ that label those $su(2)$ subalgebras. Moreover, these are classical results. To determine the properties of the Dark Monopoles at the quantum level, one could use for example semi-classical quantization.

\section{Non-abelian magnetic charge}\label{Magnetic_Charge}

One of the main properties of the Dark Monopole solution is that its magnetic field is in a direction outside the Cartan subalgebra $\mathcal{H}$. Thus, as we mentioned before, this monopole has vanishing abelian magnetic charge, since $\text{Tr}(B^{i}\phi)=0$. However, from eq.\eqref{eq:asymptotic magnetic field} we see that far from the monopole core it has a non-abelian magnetic flux in the direction $g(\theta,\varphi)\,M_{3}\,g^{-1}(\theta,\varphi)$,
with $M_{3}$ given by eq.\eqref{eq:su(2) positive alpha}. We shall define 
\begin{equation}
\zeta(\vec{r}) = a(r)\,g(\theta,\varphi)\,M_{3}\,g^{-1}(\theta,\varphi) = a(r) \,n^{a}M_{a}\,,\label{zeta-direction}
\end{equation}
which is in the direction of the monopole non-abelian magnetic flux, where $a(r)\in\mathbb{R}$ is a radial function such that $\zeta$ is regular everywhere. This implies that when $r \to 0$, $a(r)\sim r$. On the other hand, when $r\gg R_{core}$, we consider that $a(r)=1$. Then, using the fact that in this asymptotic region the gauge and the scalar fields assume the form \eqref{eq:asymptotic gauge field in Cartesian coord} and \eqref{eq:asymptotic conf. scalar field}, respectively, it is easy to verify that asymptotically $\zeta$ satisfies the conditions 
\begin{align}
D_{\mu}\zeta & =0\,,\label{eq:Killing equation for W}\\
\left[\phi,\zeta\right] & =0\,.\label{eq:Killing equation for phi}
\end{align}
Recalling the infinitesimal form of a gauge transformation for $W_{\mu}$ and $\phi$, we can conclude that the asymptotic configuration of the monopole is invariant under a gauge transformation of the form $\exp(i\zeta)$. Therefore, $\zeta$ is a Killing vector which is associated to a symmetry of the asymptotic fields of the monopole. According to \cite{AbbottDeser} and \cite{BarnichBrandt}, from the existence of a Killing vector $\zeta$ for an asymptotic symmetry one can associate a conserved charge. It is interesting to note that $\zeta$ satisfies the same equations as the scalar field $\phi$ for the 't Hooft-Polyakov monopole, outside the monopole core. Therefore, in this special case $\phi$ can be identified with the Killing vector $\zeta$. Note that if we perform an arbitrary gauge transformation $U$ on the monopole's fields then, from eqs. \eqref{eq:Killing equation for W} and \eqref{eq:Killing equation for phi}, we obtain that $\zeta$ must transform as 
\begin{equation*}
\zeta\rightarrow\zeta'=U\,\zeta\,U^{-1}\,,
\end{equation*}
in order to be a Killing vector of the transformed fields.

Moreover, since $\zeta$ and $W_{i}$ take values in the $su(2)$ subalgebra formed by $M_{a}$, we can expand them as
\begin{align}
W_{i} & =W_{ia}M_{a}\,,\nonumber \\
\zeta & =\zeta_{a}M_{a}\,,\label{eq:expansion in M}\\
D_{i}\zeta & =\left(D_{i}\zeta\right)_{a}M_{a}\,,\nonumber
\end{align}
where $\left(D_{i}\zeta\right)_{a}=\partial_{i}\zeta_{a}-e\epsilon_{abc}W_{ib}\zeta_{c}$. We shall also introduce the notation $\overline{\zeta}$ for the asymptotic configuration of $\zeta$. Then, it follows from eq.\eqref{zeta-direction} that $\overline{\zeta}_{a} = n^{a}$ is a unitary vector. Note that $\overline{\zeta}_{a}^{2}=1$ defines a 2-sphere, which we will denote by $\Sigma$. 

Now, let us define a gauge-invariant magnetic current by taking a projection of $^{*}G^{\mu\nu}$ in the direction of the Killing vector $\zeta$ as 
\begin{equation}
J_{M}^{\mu}\equiv\frac{1}{|\overline{\zeta}|y}\,\partial_{\nu}\text{Tr}\left(^{*}G^{\mu\nu}\zeta\right)\,,\label{4-current}
\end{equation}
where $|\overline{\zeta}| \equiv \sqrt{\overline{\zeta}_{a}\overline{\zeta}_{a}} = 1$. Besides that, $^{*}G^{0i}=B^{i}$ and $^{*}G^{ij}=-\epsilon_{ijk}E^{k}$. The conservation of the current $J_{M}^{\mu}$ follows from its definition as a divergence of an antisymmetric tensor and from the fact that $\text{Tr}\left(B^{i}\zeta\right)$ is twice differentiable. 

Thus, the conserved non-abelian magnetic charge is 
\begin{equation}
Q_{M}=\int_{\mathbb{R}^{3}}d^{3}x\,J_{M}^{0}=\frac{1}{|\overline{\zeta}|y}\oint_{S_{\infty}^{2}}dS_{i}\,\text{Tr}\left(B^{i}\zeta\right)=-\frac{8\pi}{e}\,.\label{eq:Magnetic charge}
\end{equation}
Note that eq.\eqref{eq:Magnetic charge} is just a measure of the non-abelian flux in the normalized $\overline{\zeta}(\theta,\varphi)$ direction. Furthermore, we must emphasize that the introduction of the radial function $a(r)$ has no contribution to the magnetic charge. This artifact was introduced so that we could define a regular magnetic current for the Dark Monopole. Besides that, as pointed out by \cite{Coleman_Aspects} there is no unambiguous way to measure the charge density of a monopole. Only the total charge makes sense. 

Let us now analyze the geometric meaning of the magnetic charge \eqref{eq:Magnetic charge}. From the asymptotic condition \eqref{eq:Killing equation for W} it follows that
\begin{equation}
\frac{1}{8\pi y}\epsilon^{ijk}\oint_{S_{\infty}^{2}}dS_{i}\textrm{Tr}\left\{ \zeta\left[D_{j}\zeta,D_{k}\zeta\right]\right\} =0\,.\label{topological_invariant1}
\end{equation}
Then, from eq.\eqref{eq:expansion in M} and using vector notation, as well as the fact that $|\zeta|=1$ when $r\to\infty$, eq.\eqref{topological_invariant1} can be written as
\begin{equation}
\frac{1}{8\pi}\epsilon^{ijk}
\oint_{S_{\infty}^{2}}dS_{i}\left\{ \hat{\zeta}\cdot\left(\partial_{j}\hat{\zeta}\times\partial_{k}\hat{\zeta}\right)-e\hat{\zeta}\cdot\overrightarrow{G}_{jk}\right\} =0\,.\label{zeta_identity}
\end{equation}
Now, using eq.\eqref{Trace_for_Ms} the expression of the non-abelian magnetic charge \eqref{eq:Magnetic charge} can be written as 
\begin{equation*}
Q_{M}= 2\, \oint_{S_{\infty}^{2}}dS_{i}\overrightarrow{B}^{i}\cdot\hat{\zeta}\,,
\end{equation*}
and from eq.\eqref{zeta_identity} we conclude that 
\begin{equation*}
Q_{M}=-\frac{4\pi}{e}\,2N_{\zeta}\,,
\end{equation*}
where 
\begin{equation*}
N_{\zeta}=\frac{1}{8\pi}\epsilon^{ijk}\oint_{S_{\infty}^{2}}dS_{i}\left\{ \hat{\zeta}\cdot\left(\partial_{j}\hat{\zeta}\times\partial_{k}\hat{\zeta}\right)\right\} \,.
\end{equation*}
As it is well-known, this integral is a topological quantity which is an integer and has the geometrical interpretation \cite{ArafuneFredGoebel} which is to measure the number of times $\hat{\zeta}$ covers $\Sigma$ as $\hat{r}$ covers $S_{\infty}^{2}$ once. For our particular Dark Monopole construction, where $\bar{\zeta}^{a} = n^{a}$, $N_{\zeta}=1$. However, in principle, one could
obtain higher magnetic charges, generalizing our construction, considering for example a gauge transformation 
\begin{equation*}
g(\theta,\varphi)=\exp\left(-i\varphi kM_{3}\right)\exp\left(-i\theta M_{2}\right)\exp\left(i\varphi kM_{3}\right),\quad k\in\mathbb{Z}\,,
\end{equation*}
which would be associated to $\hat{\zeta}$ covering $\Sigma$ $k$ times as $\hat{r}$ covers $S_{\infty}^{2}$ once. 

It is important to remark that for the Dark Monopole, the magnetic charge is not the usual one (in the abelian direction), associated to the homotopy classes of the scalar field, like in the 't Hooft-Polyakov case. 

Therefore, from the results above we can conclude that the non-abelian magnetic charge of the Dark Monopole is conserved and quantized in multiples of $8\pi/e$. And even though they are associated to the trivial sector of $\Pi_{1}(G_{v})$, the conservation of $Q_{M}$ could prevent them to decay, at least classically. However, it is necessary to analyze in more detail the stability of the Dark Monopole.

\section{Discussions and Conclusion}

In this work we have obtained a general procedure to construct magnetic monopole solutions, which we call Dark Monopoles, since their magnetic field vanishes in the direction of the generator of the $U(1)_{em}$ electromagnetic field. In order to do that, we considered theories with gauge group $SU(n)$ and a scalar in the adjoint representation. But we expect that this construction can  be generalized to other gauge groups. These Dark Monopoles must exist in some Grand Unified Theories and we analyzed some of their properties for the $SU(5)$ case.  In particular, we obtained their mass range. 

We also have shown that our monopole solution has a conserved magnetic current $J^{\mu}_{M}$ in the direction of the Killing vector $\zeta$. The associated charge is quantized and it measures the number of times $\hat{\zeta}$ covers $\Sigma$ as $\hat{r}$ covers $S_{\infty}^{2}$ once.  In principle, the conservation of this  non-abelian magnetic charge could prevent the Dark Monopoles to decay. However, the stability should be analyzed in more detail in the future.
Nonetheless, in order to discuss some cosmological aspects, let us assume for a moment that our solution is indeed stable or has a reasonable lifetime.

We expect that the Dark Monopoles were created in a phase transition in the early universe by the Kibble mechanism at a temperature of the order of the unification scale, along with the standard GUT monopoles. Under some general assumptions \cite{KolbTurner} one can show that their initial abundance $n_{M}(t_{i})$ has evolved in time according to \cite{Preskill} $$\dot{n}_{M} + 3Hn_{M} = -Dn_{M}^{2}\,,$$ where $H \equiv \dot{a}/a$ is the Hubble parameter, while $a(t)$ is the scale factor in the Robertson-Walker metric. The last term is associated to the annihilation mechanism and comes from the collision term in the Boltzmann equation. This implies that the time evolution of the monopole density strongly depends on how the monopoles interact between themselves and also on how they interact with the plasma of particles in the universe. Their motion can be described as \cite{VilenkinShellard} a Brownian motion of heavy dust particles in a gas or liquid with a slight bias in their random walks caused by the interaction between monopoles and antimonopoles. But one should note that since our monopoles have a vanishing $U(1)$ magnetic charge, there may be some differences in the annihilation mechanism, such as a different mean free path $l$ and capture radius $r_{c}$ (high-temperature regime) as well as the cross-section for radiative capture (low-temperature regime). As a consequence, we expect the so called \emph{monopole-to-entropy ratio} \cite{VilenkinShellard,WeinbergBook} to be different. However, a future detailed analysis on how the monopoles interact is necessary in order to make estimates of this ratio. 

Another relevant point is that when some possible ordinary monopoles interact with the magnetic field of our galaxy, they are accelerated. And it depends on the mass of these monopoles whether they will be ejected or slightly deflected \cite{WeinbergBook}.
In any case, the acceleration of these monopoles will drain energy from the galactic field. Now, note that since our Dark Monopoles do not interact with galactic magnetic fields, they will not be accelerated and, in principle, this means that they can cluster with the galaxy. The same reasoning can be applied to magnetic fields in galactic clusters.

Now, with regard to the Dark Matter problem we expect that Dark Monopoles might contribute to part of the mass usually attributed to Dark Matter. However, in face of the inflationary scenario, we expect that this contribution might be small. One way out of this is to investigate whether it is possible that any amount of Dark Monopoles were created during the \emph{reheating} phase after inflation through energy density fluctuations. Although even if they do not have a relevant contribution to Dark Matter, they are still an interesting solution since they are a new type of GUT monopoles.

Finally, we recall that as the case of standard GUT monopoles the mass of our Dark Monopoles is set by the GUT scale, which is beyond the energy scale of particle physics experiments, and currently direct detection is unlikely. However, as we mentioned before, further analysis is needed on how our monopoles interact and this may give some hints on the way we can look for them.

\acknowledgments

M.L.Z.P.D is grateful to CNPq for financial support. M.A.C.K. is grateful to P. Goddard, G. Thompson and I.E. Cunha for discussions.

\end{document}